# The Effect of Inflated Backrest Stiffness on Shearing Loads Estimated with Articulated Total Body

Bob J. Scurlock, Ph.D., ACTAR[*], James R. Ipser, Ph.D.[*],
Paul A. Borsa, Ph.D., ATC, FACSM[#]

[*]Department of Physics, University of Florida, Gainesville, Florida
[#]Department of Applied Physiology & Kinesiology, University of Florida, Gainesville, Florida

**Introduction**

In the construction of simulations of rear-end vehicle impacts, the Articulated Total Body (ATB) software package can be a useful tool. In this article we discuss the effect of using artificially inflated values for seat-backrest stiffness in ATB simulations. We will also present methods for quickly assessing the quality of simulation results. In this connection, we will discuss the perils of using the default contact-force models that are included in ATB package releases.

**Articulated Total Body**

The reader is referred to [1] for more thorough reviews of Articulated Total Body. Briefly, ATB is a 3-dimensional physics simulation package typically used to simulate interactions of the human body with its environment. A body is composed of an ensemble of ellipsoids connected by point-like joints. The ellipsoids can interact with pre-defined "contact planes" within the simulation environment. The allowed contact forces between the ellipsoids and contact planes are entirely specified by the user. ATB software releases typically include the Generator of Body (GEBOD) executable, which allows the user to create an ATB "dummy" of nearly any height and weight to be used by the ATB executable [2]. GEBOD uses regression equations to relate body heights or weights to appropriate ellipsoid sizes, weights, and moments-of-inertia. GEBOD also specifies joint properties.

ATB is of particular interest in vehicular accident reconstruction in the context of biomechanical physics studies as they relate to internal joint-forces, as well as external contact-forces between various planes within the occupant cabin and the vehicle occupants. In ATB, the contact planes are coupled to a vehicle frame of reference, whose dynamical properties can be completely specified as a function of time, thereby allowing the user to let the vehicle contact planes move according pre-determined motion sequences. ATB can also base the vehicle motion on acceleration pulses. Vehicle motion data can be obtained from vehicle dynamics simulation packages such as EDSMAC4 [3] and ICATS [4].

After an ATB simulation is completed, one is generally left with a series of detailed log files, which will include information such as external and internal forces, accelerations, and torques on the body versus time. Once various contact and joint forces are obtained from the ATB log files, comparisons can be made against standard sources for load tolerances based on cadaver and volunteer studies. Whether or not these types of comparisons are meaningful or valid is beyond the scope of this paper. We will mention however, a reasonable proposition is that if the internal joint forces and torques experienced during an impact on a given body part are below all known tolerance values obtained from volunteer and cadaver studies, one might reasonably conclude injury is unlikely given certain assumptions. One must be use great care however, when reasoning that simulated loads that are found to be in excess of published thresholds *necessarily* imply injury was caused. For example, generally for rear-end impacts one searches through hundreds of pages of output to find the peak anterior-posterior (A-P) shearing force imparted to the single joint connecting the "Pelvis" to the "Lower Torso" ellipsoids of the virtual ATB occupant. Given the pelvic ellipsoid represents the torso below the iliocristale, this single joint then could be thought to represent the lumbar spinal region. We have found that some ATB users like to compare such peak shearing force values to published tolerance thresholds for vertebral joints derived from laboratory tests.

In the case of injuries to the thoracic and lumbar spine region, ATB does not account for the distribution of forces across all of the structures of the upper and lower torso such as the musculature, bony structures, and connective tissues. For example, load sharing across neighboring structures within the lumbar spine and pelvis due to paraspinal and erector spinae muscle groups is not simulated. In addition, because ATB is a simple lumped mass simulator, it does not account for the individual movements of the internal organs within the abdominal and thoracic cavities. In the case of rear impacts, the solid organs within the thoracic cage (e.g. liver, lungs, heart, and spleen) are not necessarily accelerated forward due to lines of action directed solely through the thoracic spine. Rather, some fraction of the mass is accelerated forward by the costal segments, which are themselves accelerated by the backrest, though there is load sharing with the thoracic spine through the costo-vertebral joints, ribs, and sternum. Additionally, there are also lines of action extending through the shoulder girdle. This can be seen in Figure 1. Similarly, the hollow and solid organs of the lower abdominal cavity (e.g., kidneys and intestines) are not solely accelerated forward by a direct lines of action extending through the lumbar spine. Rather, lines of action can be established between the backrest, the thoracolumbar fascia, and the organs of the abdominal cavity, thereby reducing the effective abdominal weight accelerated forward solely by the lumbar spine. This can be seen in Figure 2. In the Appendix, we present a simple model describing a body being accelerated forward by a linear spring force. There we see that the peak force is proportional to $\sqrt{\text{Mass}}$. This implies a reduction in the total effective mass undergoing forward acceleration reduces the peak loads borne by the spine.

**Deconstructing the .aou File**

Generally, once a simulation is completed, ATB will output a ".aou" file, which contains a wealth of information about the simulation in a human readable format. This file should be the starting point for evaluating results derived from ATB. In this section we provide a short summary of so-called output "data cards" in the .aou file that are of special interest. As examples of the data contained in the cards, we use data that was generated by an actual rear-impact ATB simulation. The simulation was created by a biomechanical expert for use in civil litigation.

After verifying the occupant's weight and size from page 2 of the .aou file, one should verify that the $\Delta v$ used in the simulation is reasonable. For example, if the acceleration versus time is explicitly specified in tabular form as an input to ATB, this table will appear in the .aou file. The acceleration pulse should be integrated over time using a package such as ROOT [5] or Microsoft Excel to verify that the resulting $\Delta v$ agrees with the alleged value from the reconstruction. Figure 3 shows an acceleration pulse read from the subject .aou file, as well as the resulting velocity versus time.

One should carefully evaluate the various contact planes defined in the "Cards D.2" section of the .aou file. Each contact plane is typically given a meaningful name by the user such as "FLOOR" or "HEAD REST". The contact planes are uniquely defined by specifying the (x,y,z) position of three vertices along the perimeters of the various planes. For a mirror-symmetric simulation about the x-z plane such as our subject rear-end accident, one can quickly get a sense of where the planes are positioned by plotting the (x, z) coordinates of the vertices in Excel (Figure 4).

The "Cards E" section of the .aou files shows the user defined force functions used by ATB. Reference [1] contains information on how the various coefficients are used to define the functions. Generally, one will find $1^{st}$ and $2^{nd}$ order polynomials as well as some tabular-form piecewise continuous functions defined in Cards E. Figure 5 shows an example force-deflection plot for a headrest obtained by our subject .aou file. The engineer used a handheld force gauge to allegedly measure the properties of the seat (though he was only able to compress each part of the seat by about one inch). Beyond one inch, the engineer had to extrapolate the seat properties. Figure 5 shows that a reasonable linear extrapolation was used to approximate the headrest force versus deflection.

Figure 6 shows the engineer's backrest force-deflection data. Here we see what we would characterize as completely unreasonable behavior. The red line shows a linear extrapolation beyond the engineer's handheld force gauge measurements. The black curve shows the actual data used in his ATB model. This plot clearly shows the model used an effective stiffness of the order 2500 lbs/in beyond 1.2 inches of deflection!

Finally, the "Cards F.1" section defines the mapping of contact forces to contact planes. Multiple functions can be used to characterize each contact plane. In the case of our example, we were particularly interested in how the seat backrest normal force-deflection functions were fully defined in light of the questionable tabular data mentioned above and shown in Figure 6. It was noted that three functions in particular were used to define the backrest contact force. The full force is given by the expression:

$$F(C, \dot{C}) = F_1(C) + F_2(C) \times F_3(\dot{C})$$

where $F$ is the total contact force normal to the contact plane, $C$ is the total deflection (compression) of an ellipsoid into the plane of the seat backrest, and $\dot{C}$ is the rate of deflection. Note that $C$ is typically taken as a positive number for compression into the seat. This force function is made of three parts. In the example, the first function, $F_1$, is defined by the force versus deflection measured by the hand-held force gauge device along with extrapolated values. (Figure 6).

$F_2$ is a second force-deflection function and is shown in Figure 7. Figure 8 shows $F_3$. This is a velocity dependent term, which can be used for example, to account inertial effects, rate dependent viscoelastic properties, or restitution effects of the contact plane.

Figure 9 shows a ROOT-based emulation of the total force function behavior for various deflection-rate scenarios. Because of the way in which $F_3$ has been defined, we see that in the quasi-static limit, the force versus deflection converges to $F_1$. As the deflection rate increases away from 0 inches per second, the force-deflection curve quickly rises to its maximum values at 5 inches per second deflection rate. The curve then decreases monotonically as the deflection rate is further increased, converging again to $F_1$ in the large $\dot{C}$ limit. The table that defines $F_3$ includes a "feature" (or bug), which causes $F_3$ to increase from 0 to 1.0 as the deflection rate increases from 0 to 5 inches per second. This results in the odd behavior that the force model is the same in both the quasi-static and fast deflection limits. The second term in the force-deflection function causes the effective stiffness of the backrest to approach 1000 lbs/in over the first 0.9 inches of deflection!

Figure 10 shows the deflection rate versus deflection for the upper torso into the backrest. Here it is seen for the Δv = 8.5 mph scenario, the upper torso achieves an initial deflection rate of 50 inches per second into the seatback at first penetration into the backrest. Note this is about 2.8 mph or about 1/3 of the full Δv. As the upper torso is decelerated by the backrest (in the vehicle frame), the deflection rate tends to 0 in/s at maximum compression of about 1.2 inches. Thus, the force-deflection function behavior of the backrest will be near the very stiff green curve shown in Figure 9. This was verified using ROOT and is shown in Figure 11, where each piece of the total normal force-deflection function on the upper torso is shown for the 8.5 mph scenario. The total expected normal force is shown by the black curve based on the deflection rate versus deflection (Figure 10) and the definitions of $F_1$, $F_2$, and $F_3$. The corresponding ATB output is superimposed in blue dashed line. The ATB output and our expectation from first principles agree well, thereby giving us confidence that the seemingly extreme force response of the engineer's ATB model is well understood.

Figure 11 demonstrates that the total force experienced by the upper torso ellipsoid in this example is of the order 1500 lbs at maximum compression. Note most of the total force imparted to the upper torso is well above the red line. The red line shows the data input by the expert engineer to supposedly characterize the seat backrest stiffness as measured by his hand held device. Indeed, by their definitions, the functions $F_2$ and $F_3$ dominate the total normal force over $F_1$ by an order of magnitude. In fact, the values input for $F_1$ for any reasonable seatback near 100 lbs/in stiffness will have very little effect on the total force function.

Upon examining the subject .aou file, we found the force functions very familiar. Generally, ATB releases come with an example simulation of a $-x$ acceleration sled test, complete with an ATB dummy seated in a sled apparatus with a thinly cushioned seat backrest and headrest. The sled is defined by a set of contact planes and their contact force functions. In this example file, one will typically find a three-part total normal force-deflection function for the backrest. In the example file, the $F_3$ term is *identical* to $F_3$ in our subject file (Figure 8). $F_2$, shown in Figure 12, is not quite identical between the two files, but is very close. The ATB sled test example uses 1000 lbs/in whereas the engineer used 900 lbs/in. Figure 13 shows $F_1$, which are dissimilar between the two as expected, as it is this function that contains the direct measurements of backrest stiffness from the engineer's force gauge.

**Reverse Engineering the ATB Study**

The above findings prompted us to study the effect of inflating backrest stiffness values. In an attempt to do so, we used the subject .aou file in order to re-create the engineer's ATB simulation, though with more "reasonable" stiffness values equal to the engineer's own $F_1$ definition (twice his measured force gauge values), while turning off the effect of the $F_2 \times F_3$ term. This brought the effective backrest stiffness down to a reasonable 100 lbs/in. We also ran simulations using identical force functions as the engineer in order to validate that we could reproduce his results. The seatback contact planes as well as the ATB dummy were adjusted to match the subject simulation. It was understood that not all aspects of the engineer's simulation could be matched due to variations in ATB version number and other subtle differences, but reasonable agreement was expected.

Our ATB simulations were initially run at Δv = 3, 5.5, 7, and 8.5 mph. The resulting peak normal contact force versus deflection is shown for the upper torso against the backrest in Figure 14. The normal contact force is shown versus Δv. Note at Δv = 8.5 mph, the peak load on the upper torso is 200% larger by using unreasonably large stiffness values. Figure 15 shows the corresponding A-P shear force versus Δv. Here we see at Δv = 8.5 mph, the A-P shear force for the lower torso – pelvis joint is increased by nearly 75% by using large stiffness values. Also worthy of note in Figure 15 is the 3 mph result for the 1000 lbs/in backrest. We see here a value near 100 lbs of A-P shear at Δv values close to average human walking speeds.

**Generalizing the ATB Results**

In order to fully understand the implied relationship between A-P shear force, seatback stiffness, and vehicle Δv, we modified our above ATB model to use seatback stiffness values of 50, 100, 150, 200, 250, 500, and 1000 lbs/in. We ran our ATB simulator for each of these values allowing Δv to span 2, 4, 6, 8, and 10 mph, thereby establishing a grid of points in the parameter space. The results of the runs are shown in Table 1.

The data points were then imported into ROOT. Figure 16 shows an interpolated 2D surface plot of the data points in Table 1. The surface plot was used to fill a 2D histogram from which 50, 100, 200, and 300 lb A-P shear force contours were extracted. This is shown in Figure 17. The iso-shear contours show a strong dependence on Δv and stiffness values below 200 lbs/in. Note for example, the ATB results show the single joint A-P shear force can exceed 200 lbs at Δv of 6 mph for backrest stiffness of 150 lbs/in, but for softer seats of 50 lbs/in, this requires Δv of 9 mph, a 50% increase in Δv.

Using our surface fit, we can arrange our results a bit differently. Figure 18 shows the factor by which the shear force is increased by using backrest stiffness of 1000 lbs/in as opposed to the lower values indicated on the y-axis. On the x-axis is the Δv. While we note the interesting result that the shear force increases for the more reasonable backrest stiffness values for Δv<5 mph (most likely just reflecting friction on the lower torso and legs), for Δv values in excess of 7 mph, we see amplification factors between 1.2 and 2. Indeed, in our subject case, where the engineer measured the backrest stiffness to actually be about 50 lbs/in, this result shows by modeling this seatback as 1000 lbs/in, he amplifies the resulting shear force by a factor of 2. In the Appendix, we present a simple rigid rod model which emulates the behavior of ATB.

**Conclusion**

In this article we have described how to interpret a typical ATB output file. We have demonstrated the danger in naively using example simulation files commonly included in ATB releases. In the context of our subject case, we have shown the relationship between backrest stiffness values and shear forces at the "lower torso" to "center torso" joint, and how this shear force value can be inflated by increasing backrest stiffness.

*Bob Scurlock, Ph. D. is a Research Associate at the University of Florida, Department of Physics and works as a consultant for the accident reconstruction and legal community. He can be reached at BobScurlockPhD@gmail.com. His website offering free ROOT software can be found at: softwareforaccidentreconstruction.com.*

*James Ipser, Ph. D. is Professor Emeritus at the University of Florida, Department of Physics. He regularly consults and provides expert opinion in the areas of vehicular accident reconstruction and biomechanical physics. He can be reached at JIpser@gmail.com. His website can be found at JIpsier.com.*

*Paul Borsa, Ph. D. is an Associate Professor at the University of Florida, Department of Applied Physiology and Kinesology. He can be reached at pborsa@hhp.ufl.edu.*

# Appendix

**A Simple Rigid Rod Model**

One can understand the general behavior of the ATB model in our subject rear-end accident in terms of a simple "rigid rod" model. Figure 19 shows such a model comprised of three masses linked together by two joints. The masses are constrained such that they can only move along the *x*-axis, thereby ignoring effects of rotation and vertical translation. The "L" shaped mass represents the car, which via the bottom substrate interacts with $m_1$ through frictional forces, and through the linear springs, interacts with the posterior portions of $m_1$, $m_2$, and $m_3$. Using Newton's 2$^{nd}$ law, the following relations can be defined:

$$m_1 a_1 = F_1 + F_{12}$$
$$m_2 a_2 = F_2 - F_{12} + F_{23}$$
$$m_3 a_3 = F_3 - F_{23} + m_{Body} g\mu$$
$$m_1 a_1 + m_2 a_2 + m_3 a_3 = \sum_i F_i + m_{Body} g\mu$$

where $F_1$, $F_2$, and $F_3$ are the force-deflection functions given by the compression of $m_1$, $m_2$, and $m_3$ into the three linear springs of constant *k*. is the coefficient-of-friction between the side inferior to $m_3$ and superior to the substrate which represents the seat cushion. $m_{Body}$ is equal to the sum of $m_1$, $m_2$, and $m_3$. Here we may think of $m_1$ as representing all body mass above the 10$^{th}$ rib (upper torso, arms, neck, and head). $m_2$ represents the mass between the 10$^{th}$ rib and iliocristale (lower torso). $m_3$ represents all mass inferior to the iliocristale (pelvis, legs, and feet). Since we do not allow for rotation in this model, $m_1$, $m_2$, and $m_3$ are constrained to move together along the *x*-axis, therefore we can make the simplifying assumption $a_1 = a_2 = a_3 = a$. With this simplification, we get:

$$\sum_i F_i + m_{Body} g\mu = (m_1 + m_2 + m_3)a$$
$$= m_{Body} a$$

or solving for *a*:

$$a = \frac{\sum_i F_i}{m_{Body}} + g\mu$$

Solving for $F_{23}$, we get:

$$F_{23} = F_3 - \left(\frac{m_3}{m_{Body}}\right)\sum_i F_i - m_3 g\mu + m_{Body} g\mu$$

Making another simplifying assumption that all three masses compress the linear springs by the same amount, we have $C_1 = C_2 = C_3 = C$. This gives:

$$F_{23} = \left\{\left(\frac{1}{3}\right) - \left(\frac{m_3}{m_{Body}}\right)\right\}\sum_i F_i + \left(1 - \frac{m_3}{m_{Body}}\right)m_{Body} g\mu$$

Generalizing the result for an arbitrary number of mass units, we have:

$$F_J = (f_c - f_M)\sum_i F_i + (1 - f_M)m_{Body} g\mu$$

where $f_c$ is the fraction of total contact force below the joint and $f_m$ is the fraction of total body mass below the joint. From this expression we see that the shear force associated with posterior contact forces is in part attributable to the distribution of weight being accelerated as well as the distribution of force causing the acceleration. Any deficiency of force needed to accelerate a large fraction of mass must be compensated for by the joint forces in order to keep the rod in ridged uniform motion.

We can solve for an upper limit on the total contact force □$F_i$, by assuming that initial contact of the backrest springs occurs once the closing-speed reaches Δv. Note for low speed impacts, this is generally an overestimate. Ignoring friction, using work-energy we can solve for the peak force:

$$\frac{1}{2}(m_1 + m_2 + m_3)\Delta v^2 = \frac{1}{2}(3k)C_{Max}^2$$

Solving for $C_{Max}$, we get:

$$C_{Max} = |\Delta v|\sqrt{m_{Body}/3k}$$

Therefore we have:

$$Max\left(\sum_i F_i\right) = 3kC_{Max} = |\Delta v|\sqrt{3k \cdot m_{Body}}$$

Finally, solving for the peak shear force, we get:

$$F_{23}^{Max} \leq \left\{\left(\frac{1}{3}\right) - \left(\frac{m_3}{m_{Body}}\right)\right\}|\Delta v|\sqrt{3k \cdot m_{Body}} + \left(1 - \frac{m_3}{m_{Body}}\right)m_{Body} g\mu$$

Here we see behavior consistent with results observed in the ATB model, where the shear force is linearly proportional to the change-in-velocity, and proportional to the square root of seatback stiffness. A C++ version of this model was run in ROOT, where the velocity versus time relation for the vehicle was derived using the exact functional form as in our subject ATB simulation. The results are shown in Figure 20. While the exact values of the resulting shear forces differ from our ATB run, this very simple model does remarkably well capturing the general behavior and can be used for rough order of magnitude estimates to cross check against ATB values.

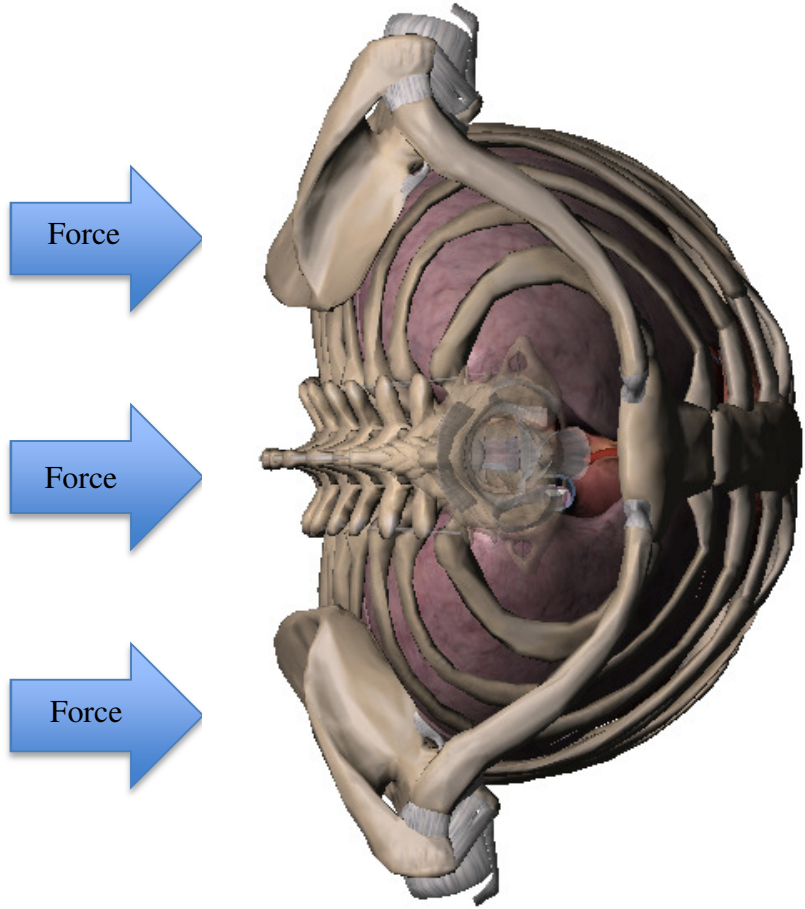

**Figure 1: View of the thoracic cage from above. Forces are imparted to the posterior portion of the body by the seat backrest during rear impact.**

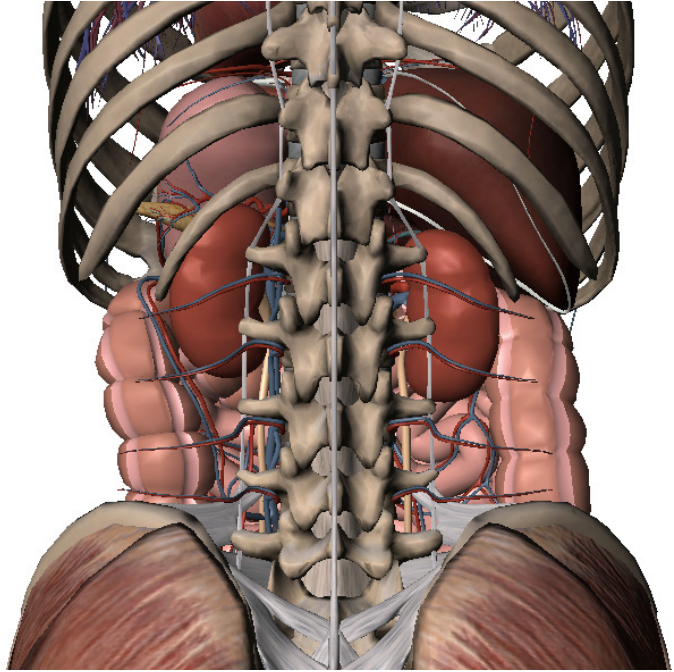

Figure 2: Posterior view of the organs within the abdominal cavity. The lumbar spine is also shown. The thoracolumbar fascia is transparent in this representation.

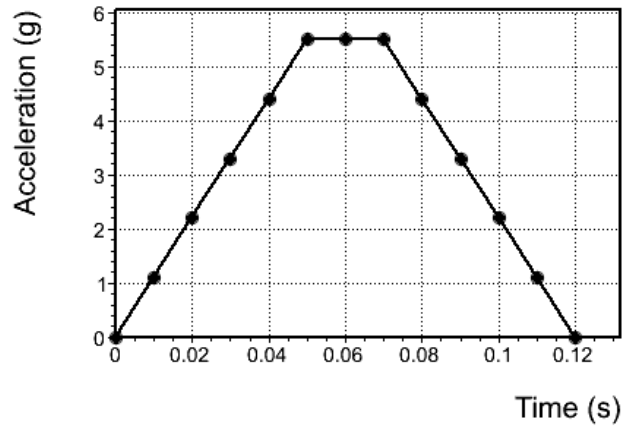
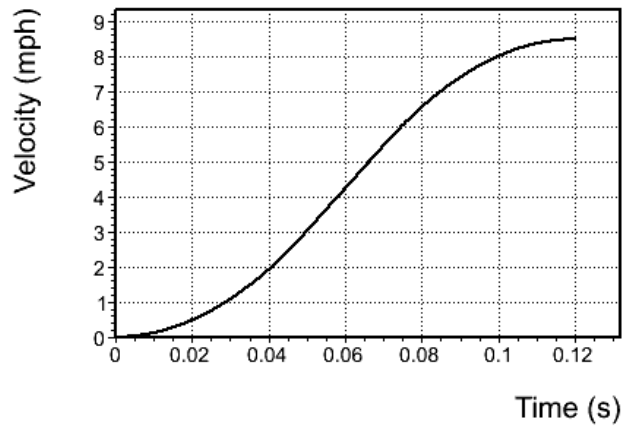

**Figure 3: Top: Acceleration pulse read from .aou file. Bottom: Velocity versus time curve obtained by integrating acceleration pulse.**

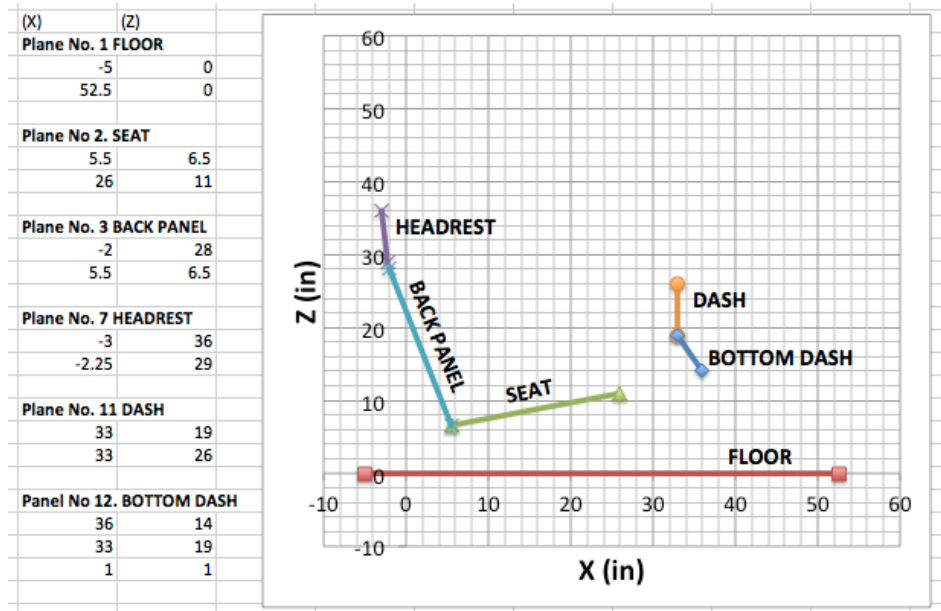

**Figure 4: (x,z) coordinates of contact plane vertices represented in Excel.**

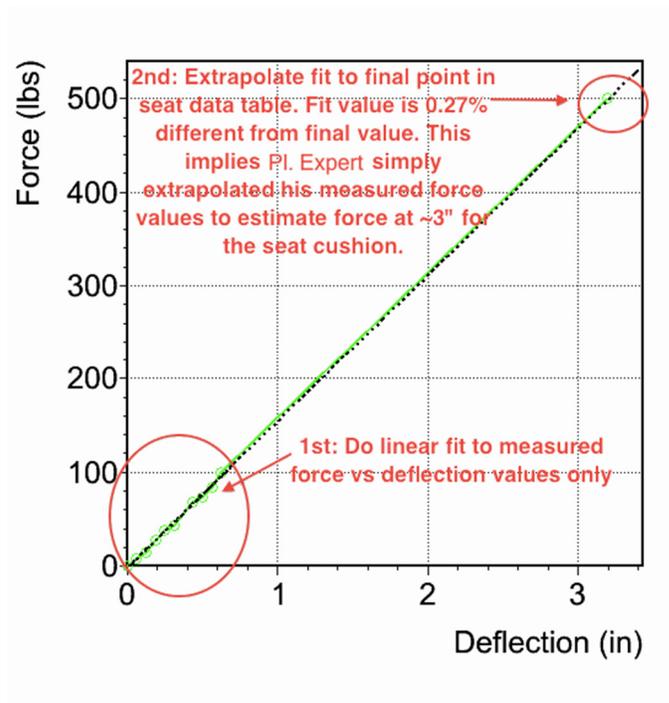

**Figure 5:** Input data for headrest contact plane. Engineer expert used a linear extrapolation to complete data table.

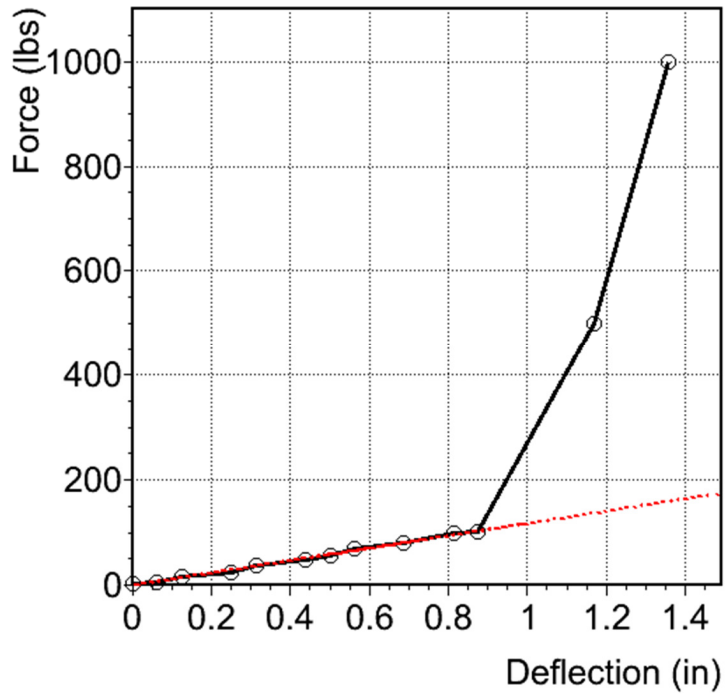

**Figure 6: (Black) Actual force versus deflection used for seat backrest contact plane. (Red) Linear extrapolation based on engineer's handheld force gauge measurements.**

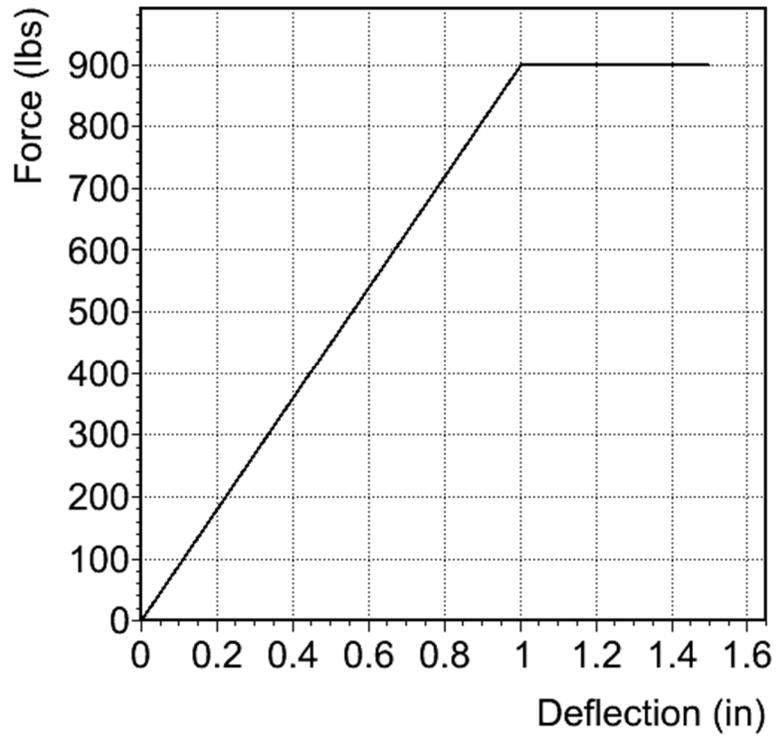

**Figure 7:** "F2" force versus deflection function which is modulated by the velocity dependent "F3".

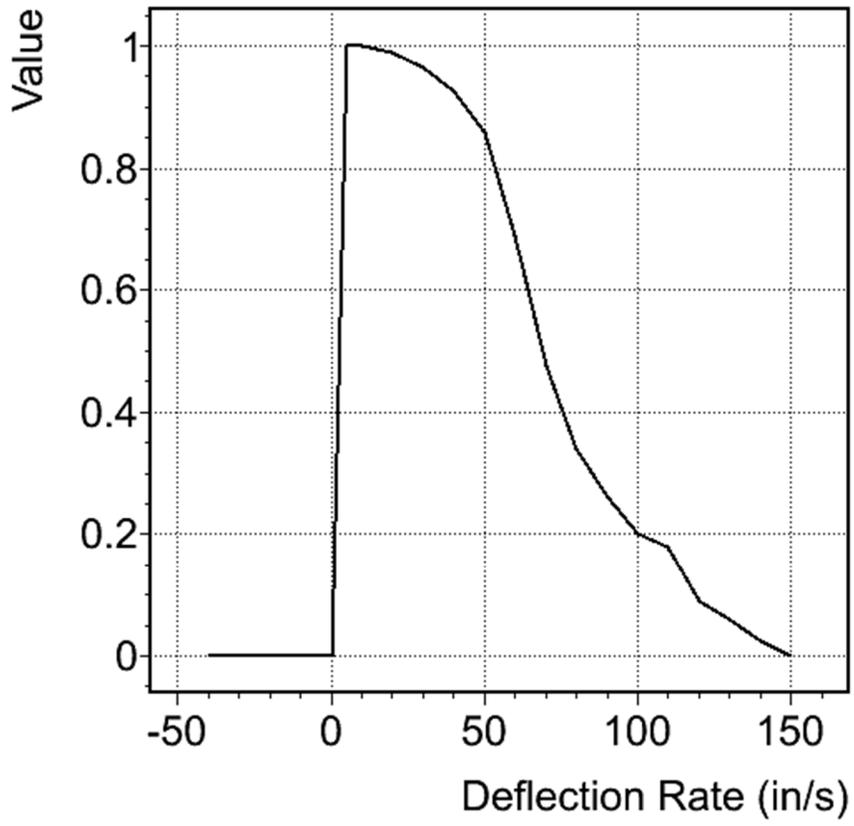

**Figure 8: "F3" velocity dependent "scale factor".**

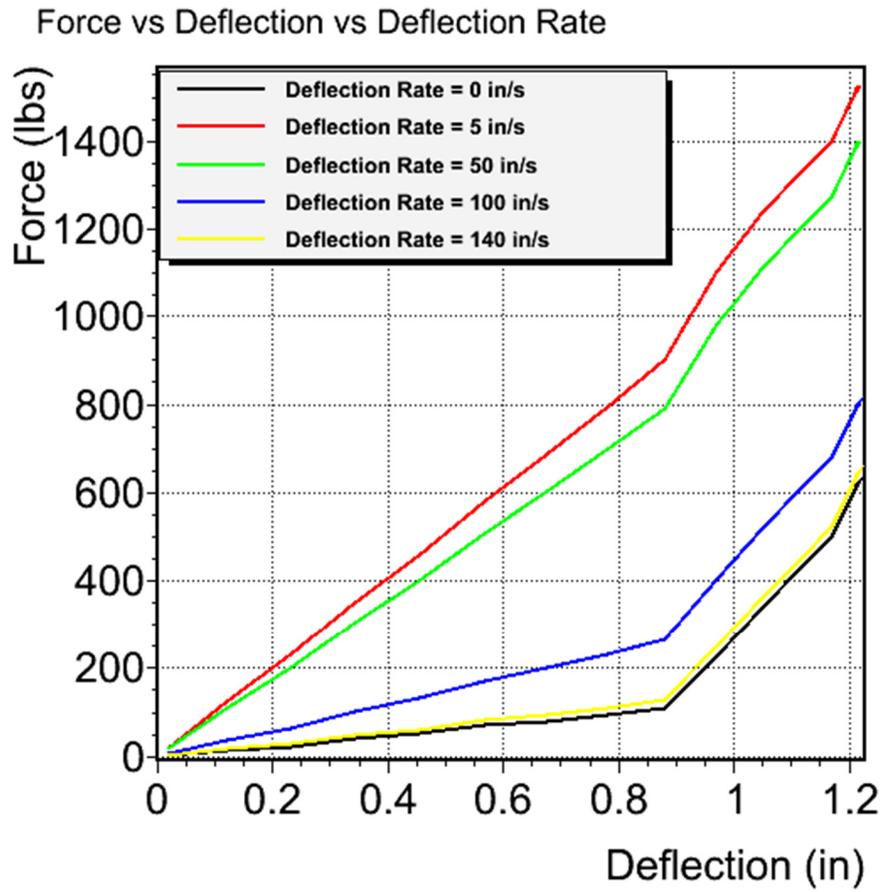

Figure 9: Total contact force function versus deflection for various deflection rate scenarios.

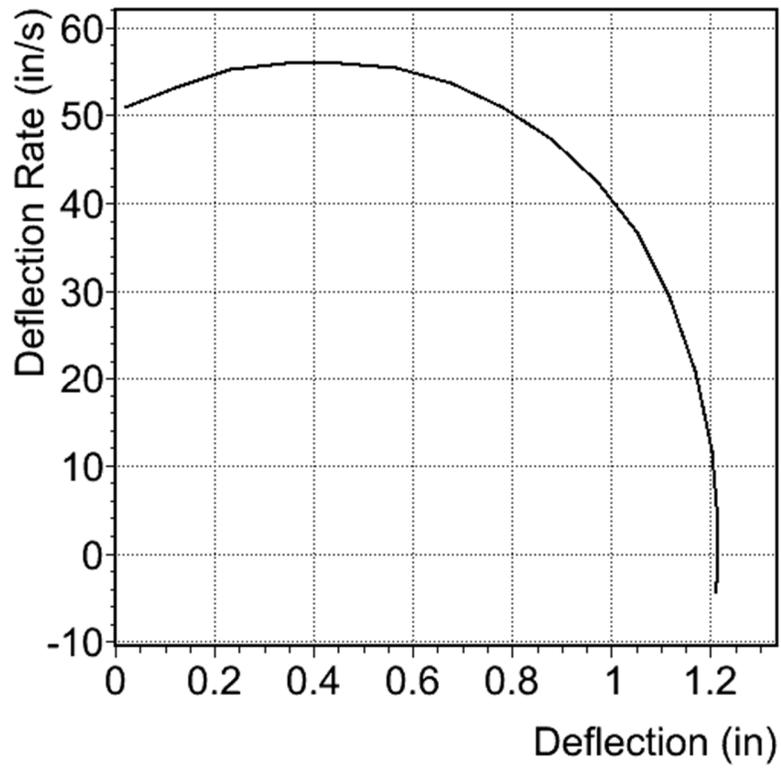

**Figure 10: Deflection rate versus deflection for upper torso ellipsoid into seat backrest contact plane.**

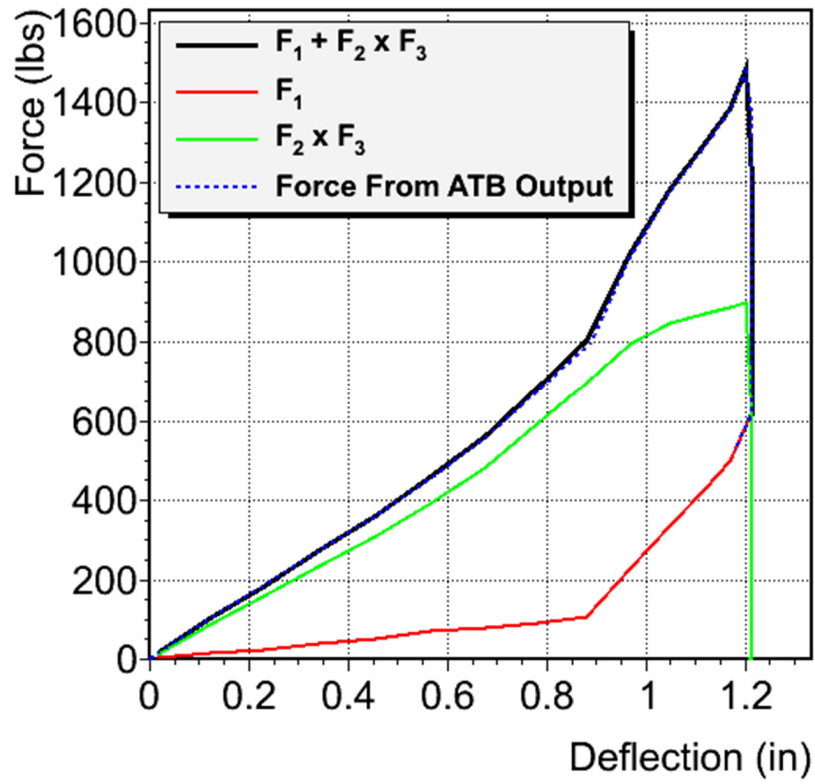

**Figure 11: Reconstruction of the total upper torso normal force from backrest. Here we compare the reported ATB output with our expectations based on the force functions.**

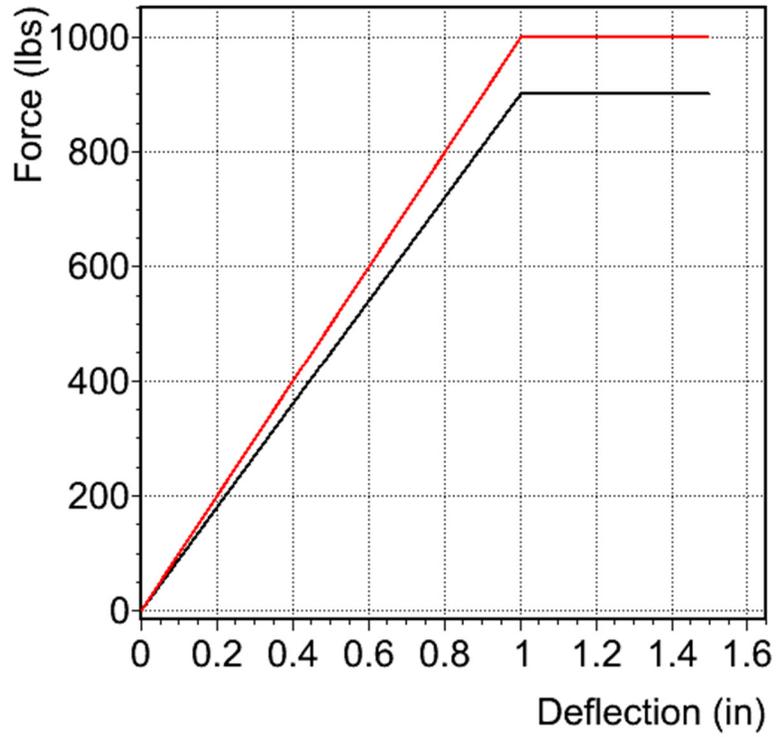

**Figure 12: (Red) F2 function found in ATB example file output. (Black) F2 function found in Plaintiff Expert's .aou file.**

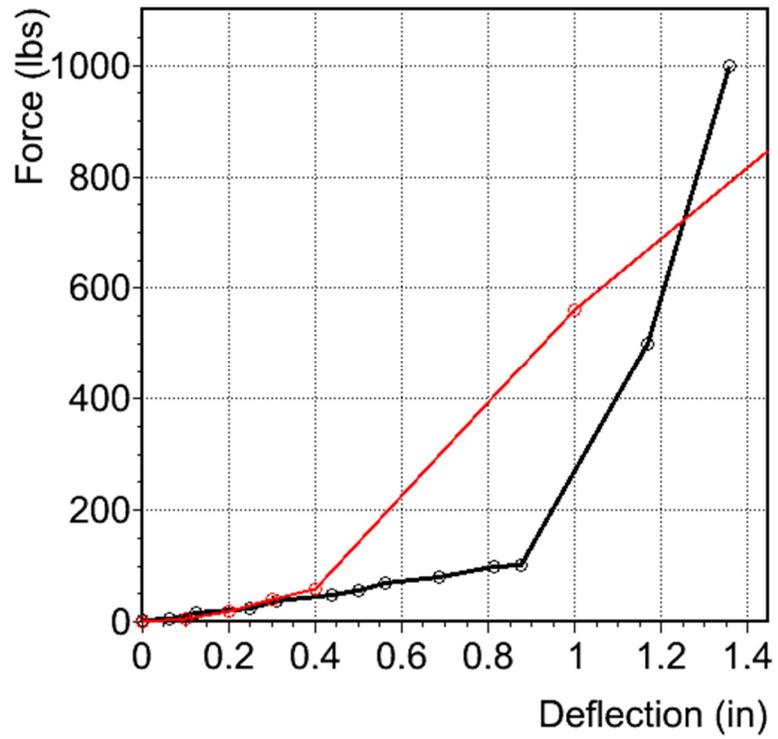

**Figure 13: (Red) F1 function found in ATB example file. (Black) F1 function used in Plaintiff expert .aou file.**

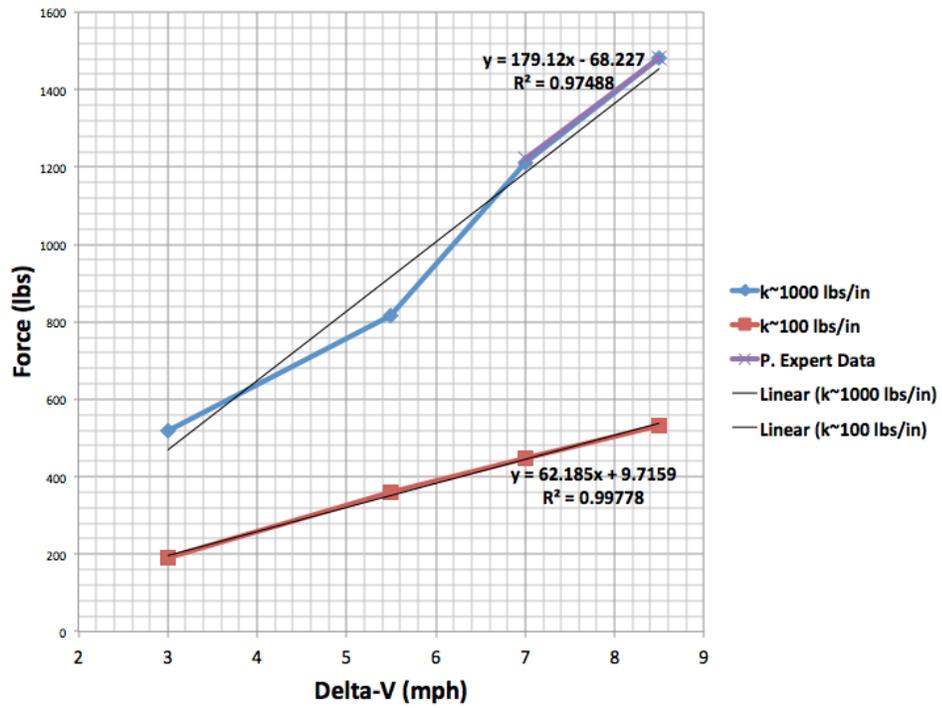

**Figure 14: Normal force versus deflection for upper torso against backrest.**

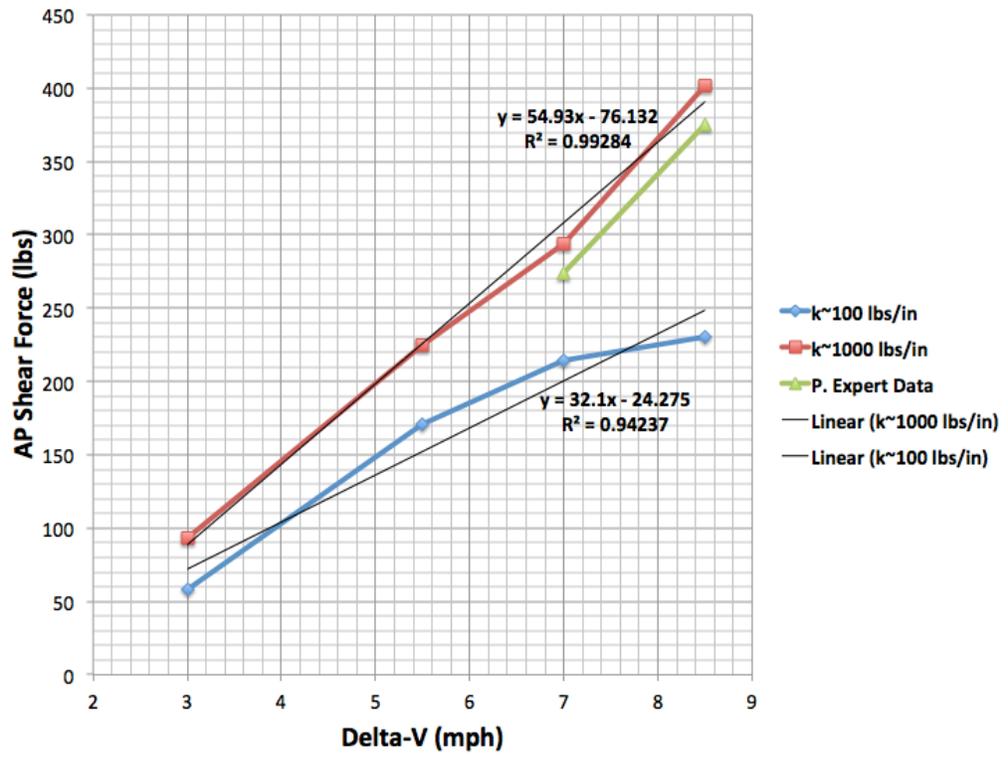

**Figure 15: A-P shear force at hip level versus delta-V.**

|  | k = 50 lbs/in | k = 75 lbs/in | k = 100 lbs/in | k = 150 lbs/in | k = 200 lbs/in | k = 250 lbs/in | k = 500 lbs/in | k = 1000 lbs/in |
|---|---|---|---|---|---|---|---|---|
| Δv = 2 mph | 55.2 | 76 | 62.8 | 53.3 | 46 | 40.1 | 45.1 | 51.4 |
| Δv = 4 mph | 82.3 | 99.8 | 107 | 111.6 | 110.6 | 106.6 | 105.2 | 89.3 |
| Δv = 6 mph | 102.2 | 146.6 | 175.9 | 198.5 | 203.7 | 208.2 | 226.4 | 235.1 |
| Δv = 8 mph | 152.1 | 155 | 201.1 | 265.8 | 290.1 | 297.6 | 338.2 | 321 |
| Δv = 10 mph | 228 | 283.5 | 205.8 | 289.7 | 352.7 | 379.6 | 405.9 | 423.5 |

**Table 1: Shear force values from ATB runs.**

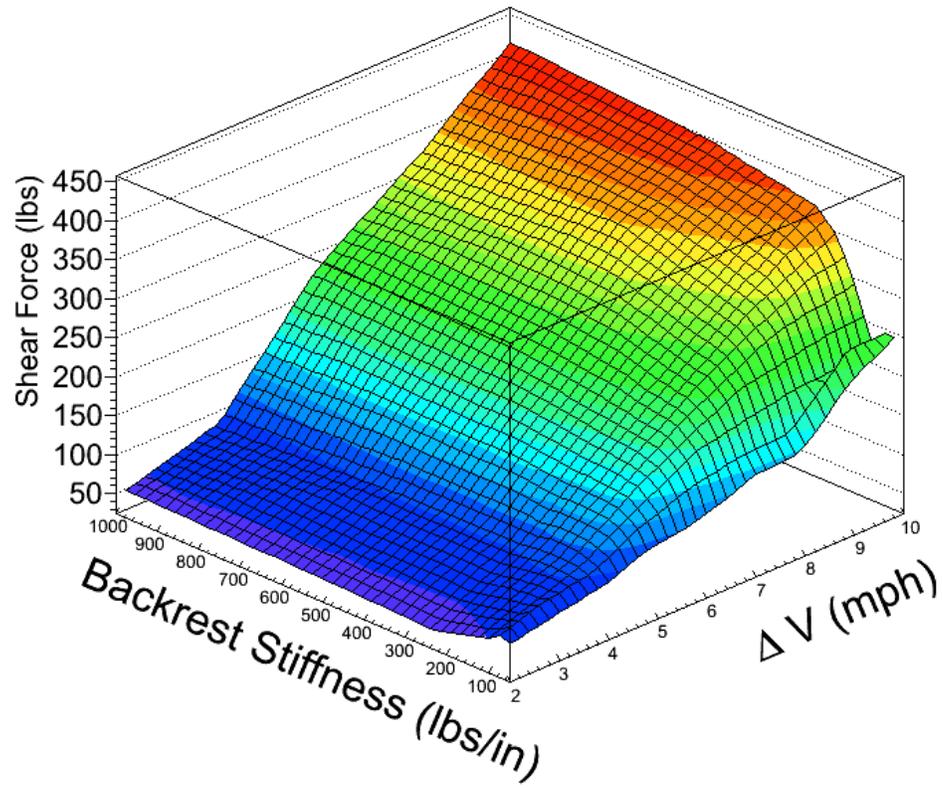

**Figure 16: Surface plot fit of ATB data.**

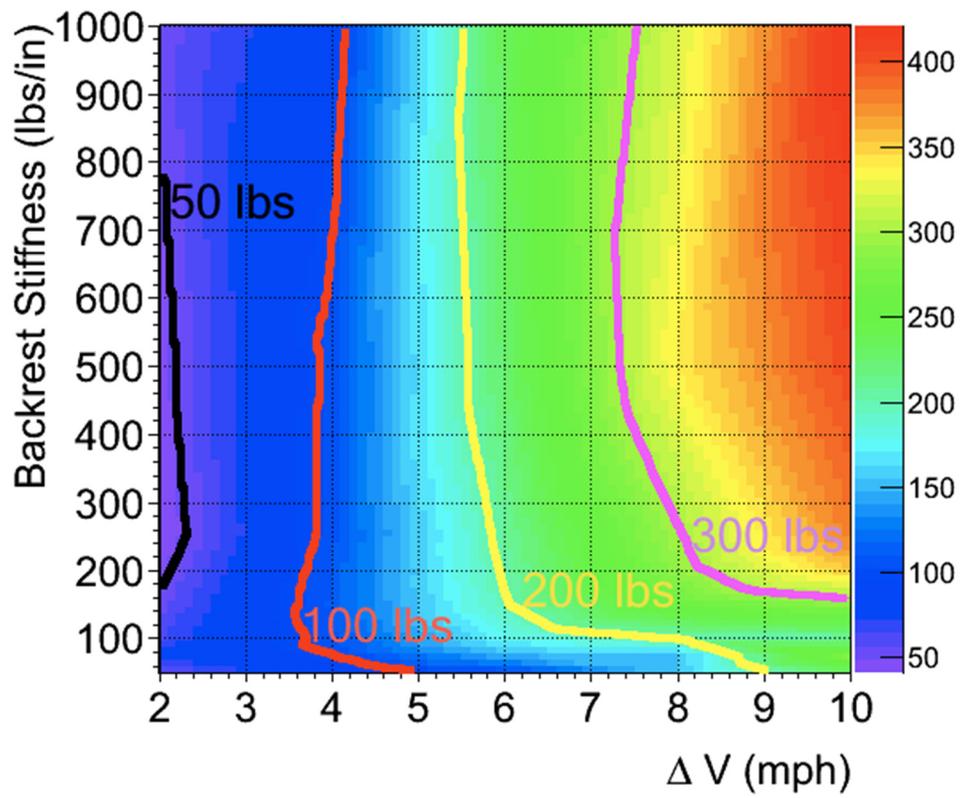

**Figure 17: 50, 100, 200, and 300 lbs A-P shear contours shown.**

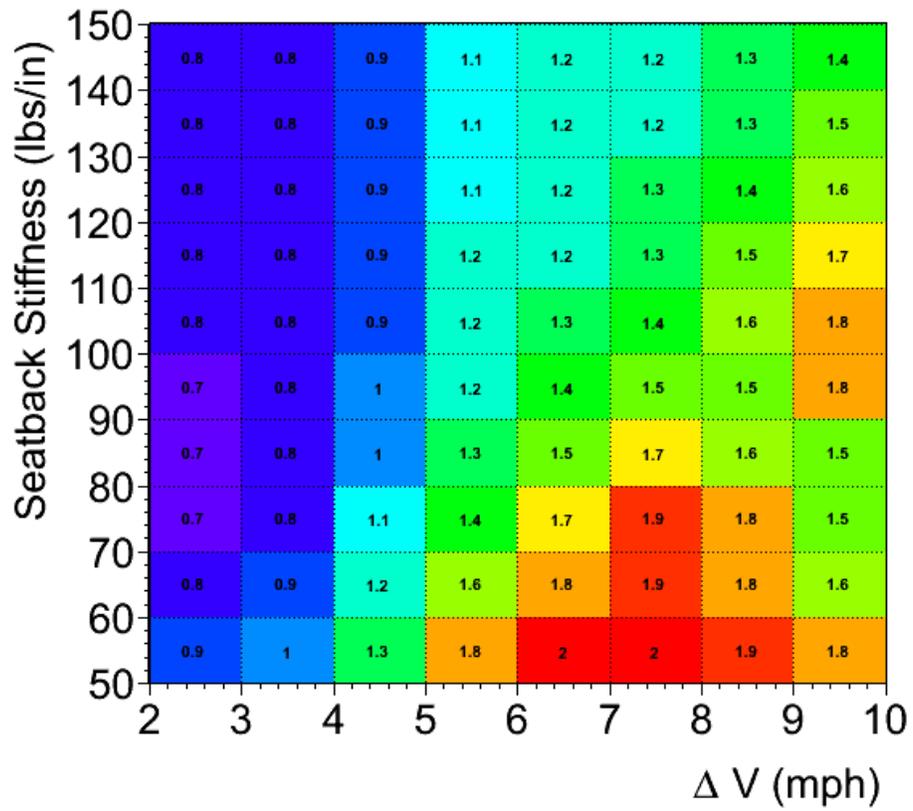

Figure 18: Amplification factors of A-P shear force which results by using 1000 lbs/in stiffness versus stiffness indicated along y-axis and Δv on x-axis.

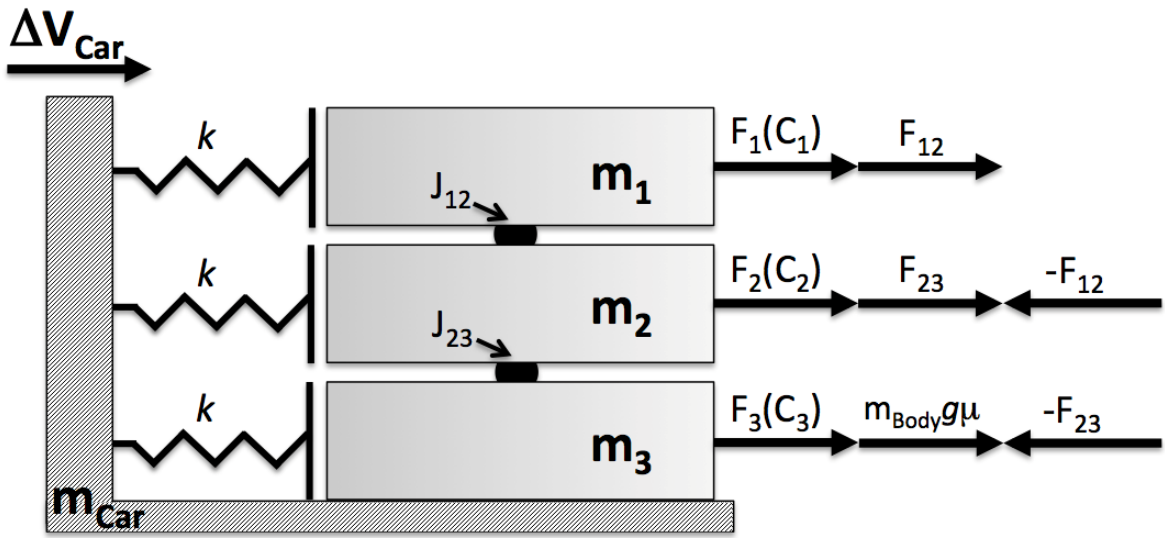

**Figure 19: Simple "rigid rod" emulator model of ATB simulation**

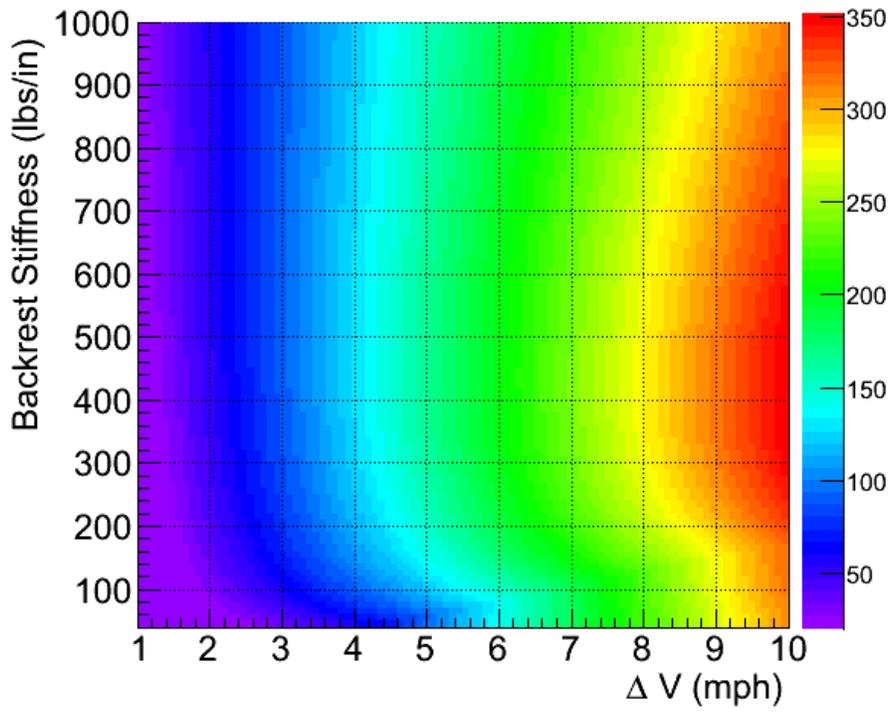

**Figure 20: A-P Shear force for rigid rod model.**